\documentclass[apjl,letterpaper]{emulateapj}
\usepackage{natbib}
\newcommand{\be}{\begin{equation}}
\newcommand{\ee}{\end{equation}}
\newcommand{\bea}{\begin{eqnarray}}
\newcommand{\eea}{\end{eqnarray}}

\newcommand{\mpci}{\mbox{$h$\,Mpc$^{-1}$}}
\newcommand{\bit}[1]{\mbox{\textbf{\emph{#1}}}}

\newcommand{\phz}{photo-\emph{z}}
\newcommand{\Phz}{Photo-\emph{z}}

\begin{document}

\title{Measuring Baryon Acoustic Oscillations with Millions of 
Supernovae}
\shorttitle{Baryon Acoustic Oscillations from Supernovae}
\author{Hu Zhan$^1$, Lifan Wang$^{2}$, Philip Pinto$^3$,
and J.~Anthony Tyson$^1$}
\shortauthors{Zhan et al.}
\affil{$^1$ Department of Physics, University of California, 
Davis, CA 95616; hzhan@ucdavis.edu, tyson@physics.ucdavis.edu \\
$^2$ Department of Physics, Texas A\&M University, 
College Station, TX 77843; wang@physics.tamu.edu\\
$^3$ Department of Astronomy and Steward Observatory, 
University of Arizona, Tucson, AZ 85721; ppinto@as.arizona.edu}

\begin{abstract}
Since type Ia Supernovae (SNe) explode in galaxies, they can, in 
principle, be used as the same tracer of the large-scale structure 
as their hosts to measure baryon acoustic oscillations (BAOs).
To realize this, one must obtain
a dense integrated sampling of SNe over a large fraction of 
the sky, which may only be achievable photometrically with 
future projects such as the Large Synoptic Survey Telescope.
The advantage of SN BAOs is that SNe have more 
uniform luminosities and more accurate photometric redshifts than 
galaxies, but the disadvantage is that they are transitory and 
hard to obtain in large number at high redshift.
We find that a half-sky photometric SN survey to redshift $z = 0.8$
is able to measure the baryon signature in the SN 
spatial power spectrum. 
Although dark energy constraints from SN BAOs are weak,
they can significantly improve the results from SN luminosity 
distances of the same  data, and the combination of the two is no
longer sensitive to cosmic microwave background priors.

\end{abstract}


\section{Introduction}
Type Ia supernovae\footnote{We consider only type Ia SNe in this
\emph{Letter}.} (SNe) have become a mature tool for studying
the cosmic expansion history 
\citep[e.g.,][]{phillips93,riess98, perlmutter99a}.
A number of SN surveys, such as  
the Sloan Digital Sky Survey (SDSS) II \citep{frieman08},
the Supernova Legacy Survey \citep{astier06},
and the ESSENCE Supernova Survey \citep{miknaitis07}
are being carried out to improve the statistics and our 
understanding of the systematics. Moreover, the SN technique
will be an integral part of almost every proposed dark 
energy survey including the Large Synoptic Survey 
Telescope\footnote{See \url{http://www.lsst.org}.} 
\citep[LSST, see][]{tyson06} and the Joint Dark Energy Mission.

The conventional SN technique, measuring only the relative luminosity
distance, $D_{\rm L}$, 
is subject to degeneracies between cosmological parameters.
For example, the SN constraint on the dark energy equation-of-state 
(EOS, $w$) parameter $w_{\rm a}$, as defined 
by $w(z) = w_0 + w_{\rm a}z/(1+z)$, is sensitive to 
the prior on the mean curvature of the universe 
\citep*{linder05b, knox06c}. The reason is that the 
response of the relative distance to a variation in $w_{\rm a}$ 
resembles that to a variation in the mean curvature and
that the SN technique lacks the calibration of absolute distances
\citep{zhan06e}. Even for a flat universe with $w(z)\equiv w_0$, 
the SN constraint on $w_0$ can be tightened considerably if the 
matter density is known to high precision \citep{frieman03}. 
Such priors may come from other techniques, such as the cosmic microwave
background (CMB), weak lensing, and baryon acoustic oscillations
\citep*[BAOs,][]{eisenstein98, cooray01b, blake03, hu03b, linder03b, 
seo03}. It has indeed been demonstrated that the latter 
three techniques are highly complementary to the SN $D_{\rm L}$
technique \citep{frieman03, seo03, knox06c}.

Since SNe explode in galaxies, their distribution bears the 
BAO imprint as well. To measure the SN spatial power spectrum, one 
needs the angular position and redshift of each SN, not its luminosity. 
Hence, the SN BAO technique does not suffer from uncertainties in 
the SN standard candle, which constitute the largest unknown in the
$D_{\rm L}$ measurements.
Nevertheless, the narrow range of the  SN intrinsic 
luminosity reduces the effect of Malmquist-like biases
and luminosity evolution. 
The SN rate traces the mass and star formation of the host galaxies
with a time delay \citep{sullivan06}. 
This means that SNe have a different clustering 
bias than galaxies that are selected by their luminosity or color. 
Finally, SNe have rich and time-varying spectral features for 
accurate estimation of photometric redshifts (\phz{}s)
\citep*{pinto04, wang07a, wang07b}, which is helpful for measuring 
BAOs from a photometric survey.

There have been discussions of using the SN weak lensing magnification 
\citep*{cooray06} and nearby SN peculiar velocities \citep*{hannestad07} 
to probe the large-scale structure. We focus on photometric SN surveys
for BAOs in this \emph{Letter} and note in passing that the SN weak 
lensing technique is more limited by shot noise than the SN BAO
technique \citep{zhan06d, zhan06e} and that the SN peculiar 
velocity technique requires precise redshift and distance measurements.

For the BAO technique to be useful, one must survey a large volume 
at a sufficient sampling density as uniformly as possible.
Although SN events are rare, the spatial density of SNe
accumulated over several years will be 
comparable to the densities targeted for 
future spectroscopic galaxy BAO surveys. 

\section{Photometric Supernova Surveys}

We assume two \phz{} SN survey models: a shallow one 
(S20k) that covers 20,000 deg$^2$ to $z = 0.8$ for 10 years, 
and a deep one (D2k) that covers 2000 deg$^2$ to $z = 1.2$
for 5 years. The S20k data may be extracted as one of the many 
products from the proposed LSST, and a dedicated ground-based 
SN survey will likely be sufficient to produce the D2k data.
Since there is no precedent for such \phz{} SN 
BAO surveys, we have to make crude estimates for the fiducial 
survey parameters and vary them to cover a wider range.

We calculate the observer-frame SN rate
based on the rest-frame SN rates
\citep{cappellaro99,hardin00,pain02,madgwick03,tonry03,blanc04,
dahlen04,neill06}.
The resulting number of SNe per steradian per unit redshift
per year (observer frame) is roughly
\[
\frac{dn}{d\Omega dz dt} \propto 
\big(e^{3.12 z^{2.1}} - 1\big) \times \left\{
\begin{array}{lr}
1 & z \le 0.5 \\ 
e^{-(z - 0.5)^2/2a^2} & z > 0.5
\end{array} \right. ,
\]
where the first term on the r.h.s.~fits the observed SN rate at
$z \le 0.55$, and the additional exponential term with $a = 0.125$
($0.21$) cuts off the distribution at $z \sim 0.8$ ($1.2$) 
for S20k (D2k).
We take the efficiency with which S20k (D2k) will produce well-observed 
SN light-curves suitable for accurate \phz{}s from 
this distribution as 50\% (100\%).
The accumulated SN surface density is then
$\Sigma = 370$ and 980 deg$^{-2}$ for S20k (7.4 million SNe in 10
years) and D2k (2 million SNe in 5 years), respectively.

We adopt a pedagogical convention and model the \phz{} error as a
Gaussian with rms $\sigma_z=\sigma_{z{\rm0}}(1+z)$ and zero 
bias\footnote{Any known \phz{} bias can be calibrated out in advance, 
so only the uncertainty of the bias matters.} $\delta z = 0$.
We assign for S20k $\sigma_{z{\rm 0}}=0.02$ and for D2k
$\sigma_{z{\rm 0}}=0.01$, which are achievable with simple
\phz{} algorithms \citep{pinto04,wang07b}.

Table~\ref{tab:model} summarizes the SN surveys. It includes 
two additional quantities: the SN clustering bias $b$, and the cut-off
wavenumber $k_{\rm max}$ for the BAO analysis, 
which is set to reduce the impact of
nonlinear growth (see \citealt{eisenstein06}
for recovering high-$k$ information from spectroscopic surveys).

\section{Forecast Method}

We use a modified forecast tool {\sc cswab}\footnote{Available
at \url{http://hzhan.net/soft/.}}
\citep{zhan06d} to assess the cosmological constraints from 
the clustering of SNe. In summary, the SN power spectrum in the
$i$th redshift bin $P_i(\bit{k}_{\rm f})$ 
reads \citep{seo03}
\bea \label{eq:psn}
P_i(k_{{\rm f}\perp},k_{{\rm f}\parallel}) 
=&&\frac{D_{{\rm f},i}^2 H_i}{D_i^2 H_{{\rm f},i}} 
\left(1+\beta_i\mu^2\right)^2 b_i^2 G_i^2 \mathcal{P}(k) \\ && \times 
\exp\left[-(c \sigma_{z,i} k_{\parallel}/H_i)^2\right] + s_i,
\nonumber
\eea
where the subscript f denotes quantities in a reference 
cosmological model (the same as the fiducial model in the
forecast), $D$ is the angular diameter distance, 
$H$ is the Hubble parameter, $\beta$ is the redshift distortion
parameter, $\mu = k_\parallel/k$,
$G$ is the linear growth factor, $\mathcal{P}(k)$ is the matter power 
spectrum at $z = 0$, and $s=n^{-1}$ is the shot noise.
The true wavenumbers $k_\perp$ and $k_\parallel$ are related to 
the references by
$k_\perp = k_{\rm f\perp} D_{{\rm f}} / D$ and
$k_\parallel = k_{\rm f\parallel} H / H_{{\rm f}}$.
Note that for a \phz{} survey with $\sigma_z \gtrsim 0.01(1+z)$, 
the radial BAO information is essentially lost, e.g., the power 
spectrum (excluding the shot noise) at the fundamental mode of BAOs
$k_\parallel\sim 2\pi/150\,\mbox{Mpc}^{-1}$ is suppressed by a factor 
of 80 or more at $z = 0.6$, though a \phz{} rms of 
$0.003(1+z)$ would recover that information.

{\sc cswab} uses the Fisher matrix to estimate the error bounds
of the parameters \citep{tegmark97b}:
\be \label{eq:FBAO}
F_{\alpha\beta}^{\rm BAO} = \sum_{i} \frac{V_{{\rm f},i}}{2} \int
\frac{\partial \ln P_i(\bit{k}_{\rm f})}{\partial q_\alpha} 
\frac{\partial \ln P_i(\bit{k}_{\rm f})}{\partial q_\beta} 
\frac{d \bit{k}_{\rm f}}{(2 \pi)^3},
\ee
where $V_{\rm f}$
is the comoving survey volume and 
$\{q_\alpha\}$ is the parameter set, which includes $b_i$, 
$\sigma_{z,i}$, $\delta z_i$, and $s_i$ of each bin and 8
cosmological parameters: $w_0$, $w_{\rm a}$, $\omega_{\rm m}$ 
(the matter density), $\omega_{\rm b}$ (the baryon density), 
$\theta_{\rm s}$ (the angular size of the sound horizon at the last 
scattering surface), $\Omega_{\rm k}$ (the curvature parameter), 
$n_{\rm s}$ (the scalar spectral index), 
and $\Delta_{\rm R}^2$ (the 
normalization of the primordial curvature power spectrum). 
The minimum marginalized error of $q_\alpha$ is $\sigma(q_\alpha) =
(F^{-1})_{\alpha\alpha}^{1/2}$. Independent Fisher matrices are 
additive; a prior on $q_\alpha$, $\sigma_{\rm P}(q_\alpha)$, 
can be introduced via $F_{\alpha\alpha}^{\rm new} = F_{\alpha\alpha}
+\sigma_{\rm P}^{-2}(q_\alpha)$.
The fiducial values of $b_i$, $\sigma_{z,i}$, and $s_i^{-1}$
are listed in Table~\ref{tab:model}, $\delta z_i = 0$,
and ($w_0$, $w_{\rm a}$, $\omega_{\rm m}$,
$\omega_{\rm b}$, $\theta_{\rm s}$, $\Omega_{\rm k}$, $n_{\rm s}$, 
$\Delta_{\rm R}^2) = (-1$, 0, 0.127, 0.0223, 
0.596$^\circ$, 0, 0.951, $2.0\times 10^{-9}$) from the \emph{WMAP} 
3-year results \citep{spergel07}.
Unless stated otherwise, we always include fairly weak 
priors $\sigma_{\rm P}(\ln b_i) = 0.3$, 
$\sigma_{\rm P}(\ln \Delta_{\rm R}^2) = 0.2$,
$\sigma_{\rm P}(\ln \theta_{\rm s}) =  
\sigma_{\rm P}(n_{\rm s}) = 0.05$, and
$\sigma_{\rm P}(\delta z_i) = 2^{-1/2} \sigma_{\rm P}(\sigma_{z,i}) =
0.25\sigma_{z,i}$.

\begin{deluxetable}{c c c c c c c}
\tablewidth{0pt}
\tablecaption{Supernova Survey Parameters
\label{tab:model}}
\tablehead{\colhead{} & \colhead{Area} & & \colhead{$n$} &  &
 & \colhead{$k_{\rm max}$} \\
\colhead{Survey} & \colhead{deg$^2$} & \colhead{$z$} &
\colhead{$h^3\mbox{Mpc}^{-3}$} & \colhead{$\sigma_z$} & 
\colhead{$b$} & \colhead{\mpci}}
\startdata
     &       & 0.3 & $4.2\times 10^{-4}$ & 0.026 & 1.18 & 0.17 \\
S20k & 20000 & 0.5 & $6.3\times 10^{-4}$ & 0.030 & 1.30 & 0.20 \\
     &       & 0.7 & $3.4\times 10^{-4}$ & 0.034 & 1.42 & 0.24 \\
\cline{1-7} \\[-2ex]
     &       & 0.3 & $4.2\times 10^{-4}$ & 0.013 & 1.18 & 0.17 \\
     &       & 0.5 & $6.6\times 10^{-4}$ & 0.015 & 1.30 & 0.20 \\
D2k  & 2000  & 0.7 & $7.8\times 10^{-4}$ & 0.017 & 1.42 & 0.24 \\
     &       & 0.9 & $5.0\times 10^{-4}$ & 0.019 & 1.54 & 0.29 \\
     &       & 1.1 & $1.8\times 10^{-4}$ & 0.021 & 1.66 & 0.34 \\[-2ex]
\enddata
\tablecomments{The redshift is central to each bin, and the width of
each bin is $\Delta z = 0.2$. 
}
\end{deluxetable}

To show the complementarity between the SN BAO and  $D_{\rm L}$
techniques, we include the 
$D_{\rm L}$ constraints at the end of Section~\ref{sec:cons}. 
The Fisher matrix for SN $D_{\rm L}$ is 
\bea \label{eq:FDL}
F_{\alpha\beta}^{\rm D_L} &=& \frac{1}{\sigma_{\rm m}^2}\int 
n_{\rm p}(z_{\rm p}) 
\frac{\partial \bar{m}_{\rm p}(z_{\rm p})}{\partial q_\alpha}
\frac{\partial \bar{m}_{\rm p}(z_{\rm p})}{\partial q_\beta} 
dz_{\rm p} \\ \nonumber
\bar{m}_{\rm p} &=& \int 
\left[5 \log D_{\rm L}(z) + M + e_1 z + e_2 z^2\right] 
p(z | z_{\rm p}) dz,
\eea
where $\sigma_{\rm m} = 0.15$ is the scatter of the SN apparent 
magnitude, the subscript p signifies \phz{} space, $M$ is the 
SN absolute magnitude, $e_1$ and $e_2$ account for possible 
SN evolution, and $p(z|z_{\rm p})$ is the probability density 
of a SN at $z$ given its \phz{} $z_{\rm p}$. 
Following \citet{albrecht06b}, 
we impose a prior of $0.015$ on $e_1$ and $e_2$ and let $M$ float.
We also include $z < 0.2$ SNe in $F^{\rm D_L}$, because they are 
important for cosmology with $D_{\rm L}$ 
\citep{linder06b}.

\begin{deluxetable}{l c c c c c c}
\tablewidth{0pt}
\tablecaption{Marginalized $1\sigma$ Errors on Selected Cosmological 
Parameters from SN BAOs \label{tab:cons}}
\tablehead{ &  & & & \colhead{$\ln \omega_{\rm m}$} & 
\colhead{$\ln \omega_{\rm b}$} & \colhead{$\Omega_{\rm k}$} 
\\[.4ex] \cline{5-7} \\[-1.2ex]
\colhead{Surveys} & \colhead{$w_0$} & \colhead{$w_{\rm a}$} &
\colhead{$w_{\rm p}$} & \multicolumn{3}{c}{$\times10^{-2}$}}
\startdata
S20k  & 0.85 & 2.9\phn & 0.22\phn & \phn\phd23 & \phn\phd38 & 1.8\phn \\
S20k+\emph{Planck} & 0.66 & 1.8\phn & 0.12\phn & 0.83 & 0.91 & 0.55 \\
\cline{1-7} \\[-2ex]
D2k  & 0.96 & 3.0\phn & 0.19\phn & \phn\phd25 & \phn\phd42 & 2.2\phn \\
D2k+\emph{Planck} & 0.74 & 2.0\phn & 0.13\phn & 0.85 & 0.91 & 0.60 \\[-2ex]
\enddata
\tablecomments{The error of the pivot EOS $w_{\rm p}$ equals 
that of a constant EOS, and the Dark Energy Task Force figure of 
merit \citep{albrecht06b} equals 
$0.052[\sigma(w_{\rm a}) \times \sigma(w_{\rm p})]^{-1}$.
}
\end{deluxetable}

\section{Cosmological Constraints}  \label{sec:cons}
The marginalized $1\,\sigma$ errors of a subset of the cosmological 
parameters are given in Table~\ref{tab:cons}. 
The two \phz{} SN BAO surveys place rather weak constraints
on the dark energy EOS even with \emph{Planck} priors.
Nevertheless, they provide moderate constraints on the matter 
density $\omega_{\rm m}$, baryon density $\omega_{\rm b}$, and 
curvature parameter $\Omega_{\rm k}$, which are helpful for the SN 
$D_{\rm L}$ technique. The smaller area of D2k is compensated by its
greater depth and better \phz{}s, so that D2k performs nearly as
well as S20k. Spectroscopic BAO surveys of similar characteristics 
can reduce the error on $\omega_{\rm m}$, $\omega_{\rm b}$, and 
$\Omega_{\rm k}$ by a factor of 2 and significantly more on $w_0$
and $w_{\rm a}$. 

The baryon signature has been detected at $\sim 3\,\sigma$ level [and 
$\sigma(\ln\omega_{\rm m})\sim 10\%$] from SDSS
Luminous Red Galaxies, both spectroscopically
\citep{eisenstein05} and photometrically \citep{blake07,padmanabhan07}.
These detections assume a flat universe with a cosmological constant
and a fixed scalar spectral index $n_{\rm s}$. 
Under the same assumptions, S20k BAO can constrain $\omega_{\rm m}$ to 
8\% and $\omega_{\rm b}$ to 15\%. If $\omega_{\rm b}$ is fixed as well
\citep[as in][]{eisenstein05}, S20k BAO can achieve
$\sigma(\ln\omega_{\rm m})=1.5\%$. 

Theoretical uncertainties  in the redshift distortion, nonlinear 
evolution, and scale-dependent clustering bias can be important
to BAOs \citep{seo05, white05, guzik07}. 
For a simple test, we replace the linear redshift 
distortion factor $(1 + \beta \mu^2)^2$ \citep{kaiser87} 
in equation (\ref{eq:psn}) with
$1 + 2 e \beta \mu^2 + f \beta^2 \mu^4$, 
where $e=f=1$ are parameters accounting for our uncertain knowledge
of the redshift distortion \citep{scoccimarro04}. 
We take  priors $\sigma_{\rm P}(e) = 0.05$ and
$\sigma_{\rm P}(f) = 0.1$ and find a less than 1\% change to 
the results in Table~\ref{tab:cons}. 
Spectroscopic (galaxy) surveys with the same redshift distribution
and coverage will see  $w_0$ and $w_{\rm a}$ errors doubled in
this test,  because they have more information to lose.

\begin{figure}
\centering
\plotone{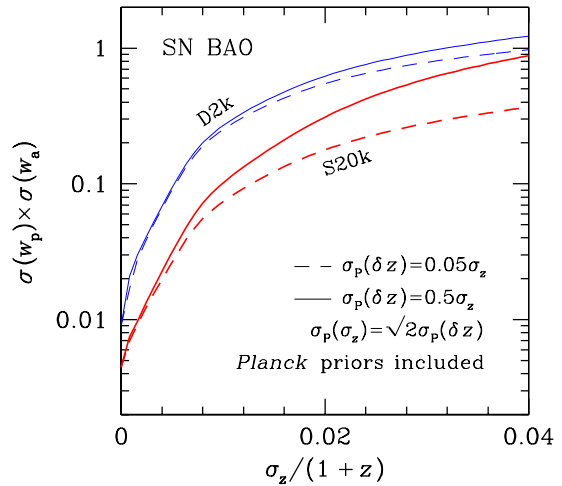}
\caption[f1]{The EP $\sigma(w_{\rm p}) \times \sigma(w_{\rm a})$ 
of S20k (thick lines) and D2k (thin lines) as a function of the rms 
\phz{} error $\sigma_z$. The priors on the \phz{} biases are 
taken to be $0.5\sigma_z$ (solid lines) and $0.05\sigma_z$ (dashed
lines), which correspond to calibrations with 4 and 400 spectra 
per redshift bin, respectively, in the Gaussian case.
To reduce the dimensions, we peg the prior on the \phz{} rms 
to that on the \phz{} bias: $\sigma_{\rm P}(\sigma_z) = 
\sqrt{2} \sigma_{\rm P}(\delta z)$. 
For comparison, LSST weak lensing, galaxy BAOs, and the two combined 
will achieve EP $\sim 0.01$, $0.02$, and $0.002$, respectively
\citep{zhan06d}.
\label{fig:conz}}
\end{figure}

\begin{figure}
\centering
\epsscale{0.93}
\plotone{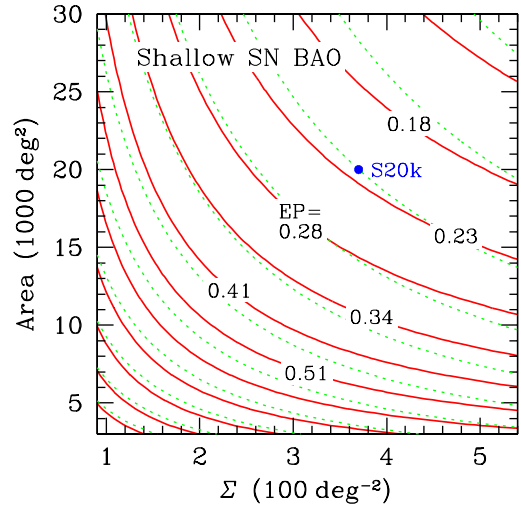}
\caption[f2]{The EP $\sigma(w_{\rm p}) \times \sigma(w_{\rm a})$ 
(solid lines) as a function of the survey area and SN surface 
density $\Sigma$ (accumulated over 10 years). 
The EP contours are spaced logarithmically. The SN distribution is 
scaled from S20k (solid circle, 7.4 million SNe) by $\Sigma$. 
\Phz{} parameters are the same as those of S20k, and the \emph{Planck} 
priors are assumed. 
The total number of SNe is held constant 
along each dotted line, which is spaced at factors of $\sqrt{2}$,
\label{fig:conn}}
\end{figure}

Figure~\ref{fig:conz} explores the dependence of the SN BAO error 
product $\sigma(w_{\rm a}) \times \sigma(w_{\rm p})$ (EP) on 
the \phz{} rms $\sigma_z$ and prior $\sigma_{\rm P}(\sigma_{z})$.
At a large $\sigma_z$, there is little radial BAO information, 
so that the Fisher matrix 
$F^{\rm BAO}$ scales roughly with the $k$-space volume, which 
is proportional to $H/c\sigma_z$. This leads to a scaling 
$\mbox{EP}\propto \sigma_z$ when $\sigma_z$ and $\delta z$ are 
known accurately (dashed lines), consistent with \citet{seo03}
and \citet{zhan06d}. When $\sigma_z \lesssim 0.008(1+z)$, 
the radial BAOs become available, and, 
hence, the EP slope steepens at smaller $\sigma_z$.

Figure~\ref{fig:conn} shows the EP contours (solid lines) 
of the shallow survey as the projected SN number 
density and survey area vary. 
Over-plotted in dotted lines are contours of the total number of SNe.
If the number of SNe is taken as a crude proxy for effort, one can 
optimize the survey (in terms of dark energy constraints) by searching
the minimum EP on the constant-effort curve. This means that for
$\Sigma \gtrsim 200\,\mbox{deg}^{-2}$ and area less than
$30,000\,\mbox{deg}^2$, one would always choose the maximum survey 
area possible for SN BAO as opposed to accumulating more SNe in a 
smaller area.

\begin{figure}
\centering
\plotone{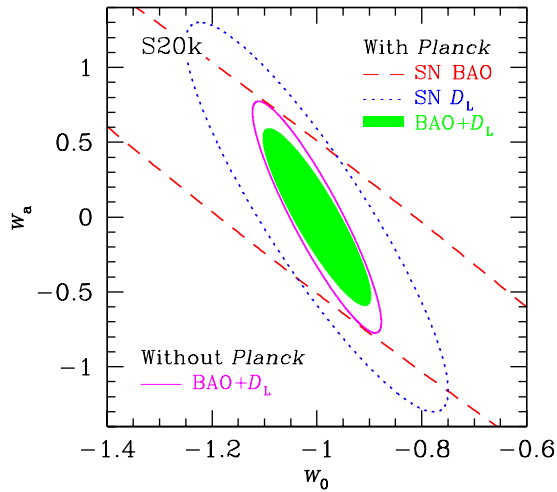}
\caption[f3]{Marginalized $1\,\sigma$ error contours of the dark energy 
EOS parameters $w_0$ and $w_{\rm a}$ from S20k SN BAOs (dashed line, 
$\mbox{EP}=0.22$ with \emph{Planck}), luminosity distances (dotted line, 
$\mbox{EP}=0.064$ with \emph{Planck}), and the two combined with 
(shaded area, $\mbox{EP}=0.012$) and without (solid line, 
$\mbox{EP}=0.018$) \emph{Planck}. 
\label{fig:snb}}
\end{figure}

Finally, we illustrate the complementarity between the SN BAO 
and  $D_{\rm L}$ techniques in Figure~\ref{fig:snb}. 
Although the S20k BAO technique (dashed line) does not place
useful constraints on dark energy, its combination (shaded area) 
with the $D_{\rm L}$ technique (dotted line) reduces the EP by a 
factor of 5.5 over the $D_{\rm L}$-alone EP. Moreover, the combined 
result (solid line) is not very sensitive to the CMB priors, because
the BAO technique can provide adequate constraints on cosmological
parameters, such as $\omega_{\rm m}$ and $\Omega_{\rm k}$, for the 
$D_{\rm L}$ technique.

\section{Discussion and Conclusions}

We have demonstrated that \phz{} SN data can be used to measure
BAOs and to constrain cosmological parameters. 
The BAO constraints on the matter density and the baryon 
density are sensitive to the priors on the curvature and the scalar 
spectral index but not to the dark energy parameters. 
The dark energy constraints from the SN BAO technique alone are not 
meaningful. However, a combination of the SN BAO and $D_{\rm L}$ 
techniques reduces the EP considerably, and the \emph{Planck} priors 
are no longer crucial. The SN BAO results in Section~\ref{sec:cons} 
are also applicable to \phz{} galaxy BAOs.
We note that
\phz{} errors are a large uncertainty for \phz{} SN cosmology.
Although our assumption about them is conservative 
compared to the results in \citet{pinto04}, further studies are 
needed to make realistic forecasts.

Long and non-uniform cadence may result in uneven sampling of 
the SN spatial distribution. 
Fortunately, LSST will be likely to always catch the SNe
(especially high-$z$ ones) at 
their maximum owing to its fast sky coverage and rapid sampling. 
Furthermore, the effect of the cadence on the SN 
depth can be simulated and determined, and methods of correcting 
for uneven depths in galaxy surveys can be applied to the SN data.

A spectroscopic SN BAO survey will be impractical, because one would 
have to revisit the sky many times spectroscopically over thousands of
square degrees to catch the SNe that occur at different times.
However, 
LSST will be able to obtain SNe in the millions over half the sky
photometrically.
This opens a window for applying the BAO technique 
to SNe and achieving more robust constraints with \phz{} SN data. 
Since this SN BAO analysis requires no additional observations than 
doing the SN $D_{\rm L}$ analysis alone, it should be a feature of 
all large-area SN cosmology analyses.
Moreover, the SN data can also help calibrate the host galaxy
\phz{}s and the \phz{} error distribution of other 
galaxies through the cross-correlation method 
\citep{schneider06,zhan06d,newman08}. 

\acknowledgments 
We thank J.~Frieman, L.~Knox, and M.~Wood-Vasey for useful 
conversations. This work was partially supported by a UC Davis 
Academic Federation Innovative Developmental Award.

\end{document}